\documentclass[aps,prd,fleqn,superscriptaddress]{revtex4}
\pdfoutput=1
\usepackage{graphicx,color,natbib}
\usepackage{amsmath,amssymb,amsfonts}

\newcommand{\bse}{\begin{subequations}}
\newcommand{\ese}{\end{subequations}}
\newcommand{\be}{\begin{equation}}
\newcommand{\ee}{\end{equation}}
\newcommand{\bea}{\begin{eqnarray}}
\newcommand{\eea}{\end{eqnarray}}
\newcommand{\ba}{\begin{array}}
\newcommand{\ea}{\end{array}}

\input amssym.def
\input amssym.tex

\usepackage[colorlinks=true, linkcolor=blue, bookmarks=true]{hyperref}

\begin{document}
\hfill%
\vbox{
    \halign{#\hfil        \cr
          IPM/P-2016/065\cr
                     }
      }
\vspace{1cm}
\title{Holographic Equilibration under External Dynamical Electric Field }
\author{M. Ali-Akbari}
\email{{\rm{m}}$_{}$ aliakbari@sbu.ac.ir}
\affiliation{Department of Physics, Shahid Beheshti University G.C., Evin, Tehran 19839, Iran}
\author{F. Charmchi}
\email{charmchi@ipm.ir}
\affiliation{School of Particles and Accelerators, Institute for Research in Fundamental
Sciences (IPM), P.O.Box 19395-5531, Tehran, Iran}

\begin{abstract}
The holographic equilibration of a far-from-equilibrium strongly coupled gauge theory is investigated. The dynamics of a probe D7-brane in an AdS-Vaidya background is studied in the presence of an external time-dependent electric field. Defining the equilibration times $t_{eq}^c$ and $t_{eq}^j$, at which condensation and current relax to their final equilibrated values, receptively, the smallness of transition time $k_M$ or $k_E$ is enough to observe a universal behaviour for re-scaled equilibration times $k_M k_E (t_{eq}^c)^{-2}$ and $k_M k_E (t_{eq}^j)^{-2}$. Moreover, regardless of the values for $k_M$ and $k_E$, $t_{eq}^c/t_{eq}^j$ also behaves universally for large enough value of the ratio of the final electric field to final temperature. Then a simple discussion of the static case reveals that $t_{eq}^c \leq t_{eq}^j$. For an out-of-equilibrium process, our numerical results show that, apart from the cases for which $k_E$ is small, the static time ordering persists. 
\end{abstract}

\maketitle

%\tableofcontents

\textit{\textbf{Introduction}}: Describing time evolution of far-from-equilibrium processes is an attractive but difficult open problem, especially at strong coupling. An important consequence of the AdS/CFT correspondence, or more generally gauge-gravity duality, is that out-of-equilibrium dynamics in strongly coupled gauge theory is dual to a time-dependent problem in a classical gravity \cite{Maldacena:1997re, CasalderreySolana:2011us}. This new theoretical tool is applied to investigate various aspects of the far-from-equilibrium physics. Tracing the (instantaneous)thermalization \cite{Chesler:2008hg} and isotropization \cite{Heller:2013oxa} processes of quark-gluon plasma is one of the significant aspects which has attracted a lot of attention in recent times. Moreover, studying quantum quench, which is also a far-from-equilibrium situation, is another motivation to use the gauge-gravity duality \cite{Das:2011nk}.

D3-D7 system is a reasonable candidate to explain different properties of QCD-like theories \cite{Erdmenger:2007cm}. In fact, the fundamental matter (quark) is added to the gauge theory via introduction of $N_f$ probe D7-branes in the background sourced by $N_c$ colour D3-branes which is normally $AdS_5\times S^5$ or AdS-black hole geometry \cite{Karch:2002sh}.  By probe (limit) we mean the back-reaction of D7-branes on the background can be ignored. As a result, DBI action describes the dynamics of the D3-D7 sytem in the probe limit \footnote{Notice that the CS action does not contribute for the case we consider in this paper.}. This model, apart from the static case, is applied to understand non-equilibrium processes which happen on the probe D7-brane such as black hole formation \cite{Das:2010yw, Hashimoto:2010wv}, time-dependent baryon injection \cite{Hashimoto:2010wv}, Schwinger effect \cite{Hashimoto:2014dza} and time-dependent meson melting \cite{Ishii:2014paa, Ali-Akbari:2015ooa}.

In the time evolution of non-equilibrium process, discussing various time-scales of relaxation is an interesting issue. In the context of gauge-gravity duality for a strongly coupled gauge theory with a holographic dual different time-scales has been introduced and studied in the literature. Furthermore time ordering of these time-scales is also addressed in more recent papers such as \cite{Ali-Akbari:2016sms, Attems:2016ugt}. Understanding how a non-equilibrium system relaxes to its final equilibrated state and what quantities are more effective during the relaxation are interesting problems to investigate. In order to do so, we consider a far-from-equilibrium strongly coupled gauge (${\cal{N}}=2$ super Yang-Mills) theory which is dual to a probe D7-brane in the AdS-Vaidya background. D7-brane is needed for introducing external time-dependent electric field and the AdS-Vaidya background resembles thermalization on the gauge theory side. Therefore, we heat up the gauge theory, subject to an external time-dependent electric field. Thus energy injection into the system is done by increasing external electric field and temperature. In other words, there are two ways where the system at hand deviates from its equilibrium. Moreover there are two parameters, $k_M$ and $k_E$, which are describing the rate of the energy injection. The problem we are interested in is how the relaxation of the system is affected by $k_M$, $k_E$ and the final values of the temperature and electric field.

In the following we use the same notations as ones introduced in \cite{Ishii:2014paa} and refer the reader to \cite{Ishii:2014paa, Ali-Akbari:2015ooa, Ali-Akbari:2015bha, Ishii:2015qmj, Hashimoto:2014yza} for more details.

\textbf{\textit{Model}}: AdS-Vaidya background is considered as a toy model of describing thermalization process in the context of the gauge-gravity duality. This background specifies black hole formation in the gravity theory which corresponds to thermalization process in the dual gauge theory (in the gluon sector). The metric of AdS-Vaidya background is given by
\be%
ds^2 =G_{MN} dx^M dx^N= \frac{1}{z^2} \left[-F(V,z) dV^2 - 2 dV dz + d{\vec{x}}_3^2 \right]
+ d\varphi^2+\cos^2\varphi ~d\Omega_3^2+\sin^2\varphi ~d\psi^2,
\ee%
where $F(V,z) = 1-M(V) z^4$. We set the radius of AdS space-time to be one. The above metric is written in Eddington-Finkelstein coordinates where the radial direction is represented by $z$ and $V$ shows the null direction. The boundary, where the gauge theory lives, is at $z=0$ and $V$ is the time coordinate on the boundary. The arbitrary function $M(V)$, related to the temperature of the gauge theory, represents the mass of the black hole which changes as time passes by until it reaches a constant value.

In order to introduce a dynamical electric field in the gauge theory, we need to add a probe brane to the bulk. The dynamics of the D7-brane is explained by the DBI action
\be\label{action} %
 S=-\tau_7\int d^8 \xi\sqrt{-\det(g_{ab}+2\pi\alpha' F_{ab})}~,
\ee %
where tension of the D7-brane is denoted by $\tau_7$ and $a,\, b$ are the brane coordinates. $g_{ab} = G_{MN} \partial_a x^M  \partial_b x^N$ is the induced metric and $F_{ab}$ is the gauge field strength on the brane. The D7-brane is embedded along the six directions of the bulk metric, $\vec{x}$ and $\Omega_3$. We choose the other two coordinates on the brane to be null coordinates $u$ and $v$. For the rest of the bulk coordinates we choose the ansatz
\be\label{a1} %
 V=V(u,v),\ \ z=Z(u,v),\ \ \varphi=\Phi(u,v),\ \ \psi=0.
\ee %
Since we are interested in studying the effect of the external time-dependent electric field on the equilibration time-scales, we also adopt an ansatz for the gauge filed as below $(x\equiv x_1)$
\be\label{a2}%
A_x(V,Z)=a_0(V,Z)+a_x(V,Z)=\int^V_{V_0} E(V')dV'+a_x(V,Z),
\ee%
and all other components of the gauge field are set to zero. $V_0$ is an initial time reference which can be set, for instance, equal to zero or $-\infty$.
Now by substituting \eqref{a1} and \eqref{a2} in the DBI action \eqref{action}, it is easy, but lengthy, to find the following equations of motion
\bse\label{eom}\begin{align}
V_{,uv}=&\frac{Z^3}{2} {a_x}_{,u} {a_x}_{,v}
+\frac{1}{2}\left(F_{,Z}-\frac{5F}{Z}\right)V_{,u}V_{,v}
+\frac{3}{2}\, \textmd{tan}(Z\Psi)\Big((Z\Psi)_{,u}V_{,v}+(Z\Psi)_{,v}V_{,u}\Big)
+\frac{3}{2}\,Z(Z\Psi)_{,u}(Z\Psi)_{,v}\,,\\
Z_{,uv}=&-\frac{FZ^3}{2} {a_x}_{,u} {a_x}_{,v}
-\frac{F_{,V}}{2}\,V_{,u}V_{,v}
+\frac{1}{2}\left(\frac{5F}{Z}-F_{,Z}\right)
\left(FV_{,u}V_{,v}+V_{,u}Z_{,v}+V_{,v}Z_{,u}\right)\nonumber\\
+&\frac{3}{2}\,\textmd{tan}(Z\Psi)\Big((Z\Psi)_{,u}V_{,v}+(Z\Psi)_{,v}V_{,u}\Big)
-\frac{3F}{2}\,Z(Z\Psi)_{,u}(Z\Psi)_{,v}
+\frac{5}{Z}\,Z_{,u}Z_{,v}\,,\\
\Psi_{,uv}=
&\frac{Z^2\Psi}{2}\Big(F-\frac{3\,\textmd{tan}(Z\Psi)}{Z\Psi}\Big){a_x}_{,u} {a_x}_{,v}
+\frac{\Psi}{2Z}\left(F_{,Z}-\frac{5F}{Z}+\frac{3\textmd{tan}(Z\Psi)}{Z^2\Psi}\right)
\left(FV_{,u}V_{,v}+V_{,u}Z_{,v}+V_{,v}Z_{,u}\right)\nonumber\\
+&\frac{\Psi F_{,V}}{2Z}\,V_{,u}V_{,v}
+\frac{1}{2Z^2}\Big(1-3\,Z\Psi\,\textmd{tan}(Z\Psi)\Big)
\Big((Z\Psi)_{,u}Z_{,v}+(Z\Psi)_{,v}Z_{,u}\Big)\nonumber\\
+&\frac{3}{2}\left(F\Psi+\frac{\textmd{tan}(Z\Psi)}{Z}\right)\,(Z\Psi)_{,u}(Z\Psi)_{,v}
-\frac{3\Psi}{Z^2}\,Z_{,u}Z_{,v}\,,\\
{a_x}_{,uv}=&\frac{1}{2Z}\Big(Z_{,u}\,{a_x}_{,v}+Z_{,v}\,{a_x}_{,u}\Big)
+\frac{3}{2}\,\textmd{tan}(Z\Psi)
\Big((Z\Psi)_{,u}\,{a_x}_{,v}+(Z\Psi)_{,v}\,{a_x}_{,u}\Big),
\end{align}\ese
where $\Psi(u,v)\equiv\frac{\Phi(u,v)}{Z(u,v)}$. Moreover, in order to solve the above equations, we need to specify the boundary and initial conditions. These conditions are similar to the ones considered in \cite{Ishii:2014paa, Ali-Akbari:2015bha, Hashimoto:2014yza}. Here we do not repeat the details of calculations and only state the final results. Therefore, boundary conditions are given by
\begin{align} %
u=&v:~Z=0,~\Psi=m,~V_{,v}=2Z_{,u},~a_{x}=a_0(V),\nonumber\\
u=&v+\frac{\pi}{2}:~(Z\Psi)=\frac{\pi}{2},~Z_{,u}=Z_{,v},~V_{,u}=V_{,v},~
{a_{x}}_{,u}={a_{x}}_{,v},
\end{align}
and for initial conditions, we have
\begin{align}
&V(u,0)=m^{-1}\big(\phi(u)-\sin\phi(u)\big)+V_{\textrm{ini}},
~~~~~~Z(u,0)=m^{-1}\sin\phi(u),\nonumber\\
&\Psi(u,0)=\frac{m\phi(u)}{\sin\phi(u)},
~~~~~~~~~~~~~~~~~~~~~~~~~~~~~~~~{a_{x}}(u,0)=0,
\end{align}
where $V_{\textrm{ini}}\le 0$ and $\phi(u)$ is an arbitrary function corresponding to the residual coordinate freedom on the
initial surface and $V_{\textrm{ini}}$ is an integration constant.

In this paper the functions for $M(V)$ and $E(V)$ that we will work with are
\bea %
 M(V)=  \left\{%
\begin{array}{ll}
    0 & V<0, \\
    \frac{M_f}{k_M}\left[V-\frac{k_M}{2\pi}\sin(\frac{2\pi V}{k_M})\right] & 0 \leqslant V \leqslant k_M, \\
    M_f & V>k_M ,\\
\end{array}%
\right.
\hspace{5 mm}
% \eea %
% and
% \bea %
 E(V)= \left\{%
\begin{array}{ll}
    0 & V<0, \\
    \frac{E_f}{k_E}\left[V-\frac{k_E}{2\pi}\sin(\frac{2\pi V}{k_E})\right] & 0 \leqslant V \leqslant k_E, \\
    E_f & V>k_E ,\\
\end{array}%
\right.
 \eea %
where $k_M$ ($k_E$) are the time interval in which the mass of the black hole (external electric field) increases from zero to $M_f$ $(E_f)$ which is constant. Note that the radius of the event horizon is $r_h=M_f^{\frac{1}{4}}$.

According to the gauge-gravity duality, our observables in the gauge theory can be obtained by near boundary expansion of the corresponding fields on the gravity side. In the case at hand, the observables are time-dependent condensation $c(V)$ and current $j(V)$. These quantities are respectively related to the near boundary expansion of $\Psi(u,v)$ and $a(V,Z)$ as follows \cite{Hoyos:2011us, Hashimoto:2014yza}
\bse\begin{align}
\Psi(V,Z)&=m+\left(c(V)+\frac{m^3}{6}Z^2\right)+...,\\
a_x(V,Z)&=\dot{a}_0(V)Z+\frac{1}{2}j(V)Z^2+...\ .
\end{align}\ese
In order to determine suitable criteria for equilibration of the system, we define the following time-dependent parameters
\be\begin{split}
\epsilon^c(V)&=\frac{|c(V)-c_{eq}|}{c_{eq}},\cr
\epsilon^j(V)&=\frac{|j(V)-j_{eq}|}{j_{eq}},
\end{split}\ee
where $c_{eq}$ and $j_{eq}$ are the final equilibrated values of the condensation and current, respectively. These values are found from our numerical calculations. Then equilibration time for condensation (current) is defined as the time at which $|\epsilon^c(V_{eq})|<0.03$ ($|\epsilon^j(V_{eq})|<0.03$) and stays below this limit afterwards.

We continue our discussion for the case $m=1$ and $V_0=0$ without loss of generality. Then condensation $c(V)$ can be numerically found by solving the equations of motion \eqref{eom}. Since the shape of the D7-brane in an arbitrary background is described by $\Psi(V,Z)$ \cite{CasalderreySolana:2011us}, when the condensation reaches its final equilibrated value the shape of the brane is fixed and does not change anymore. However, from the static case, we know that the static current depends on the electric field as well as the shape of the brane in the probe limit. Therefore one may speculate that the equilibration time for the condensation is always smaller than or equal to the equilibration time for the current, that is $t_{eq}^c\leq t_{eq}^j$. We will check this naive statement in the next section.

\begin{figure}[ht]
\begin{center}
\includegraphics[width=8 cm]{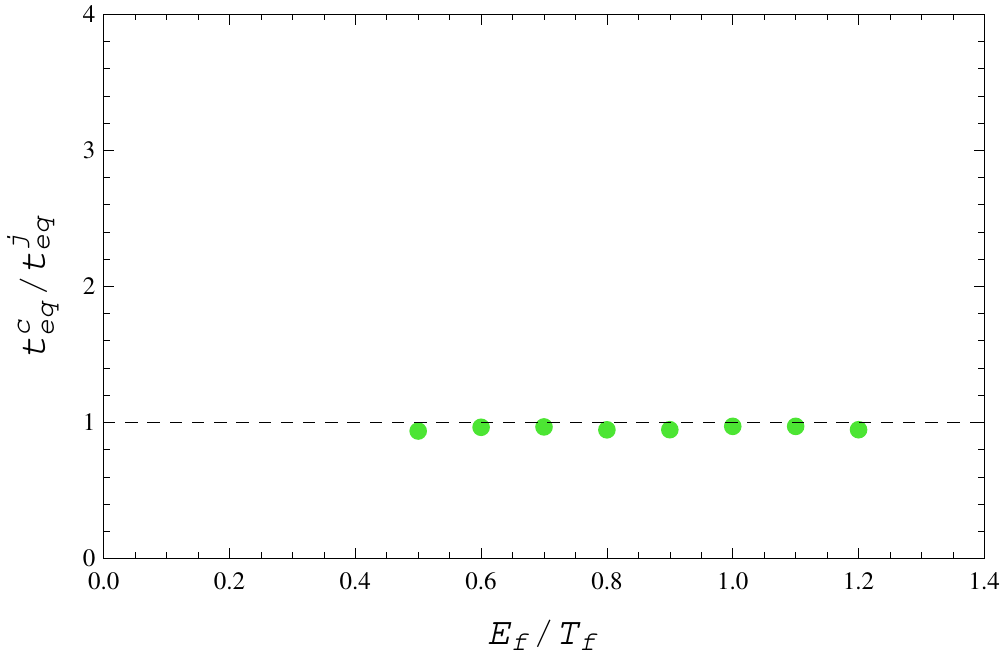}
\hspace{0.6 cm}
\includegraphics[width=8 cm]{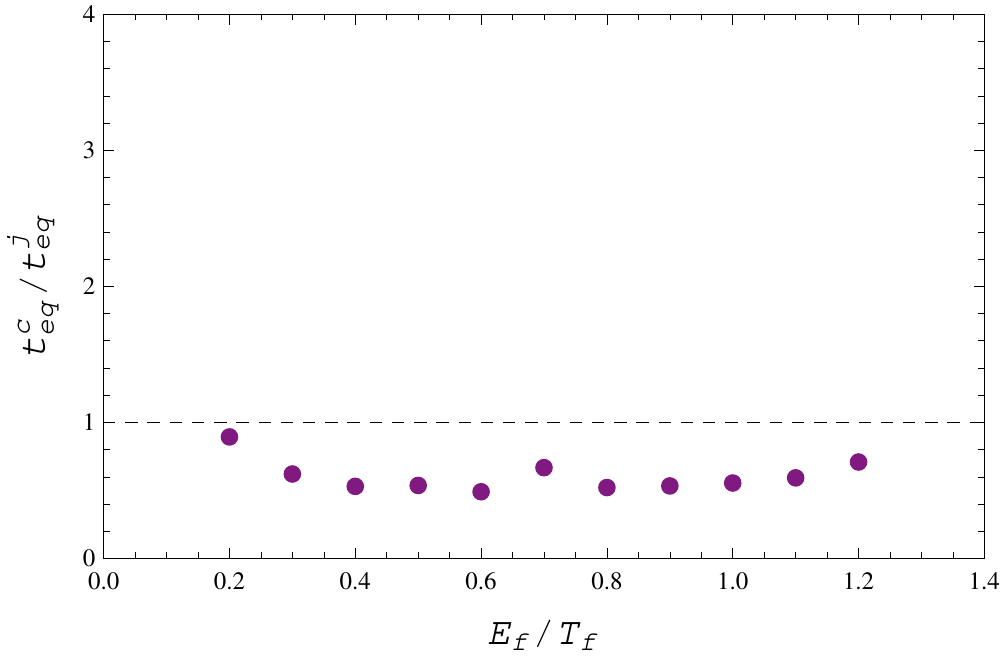}
\caption{$t^c_{eq}/t^j_{eq}$ in terms of $E_f/T_f$ for slow-slow(left) and slow-fast(right) transitions for $T_f=5$. \label{slowfast} }
\end{center}
\end{figure}%
\begin{figure}[ht]
\begin{center}
\includegraphics[width=8 cm]{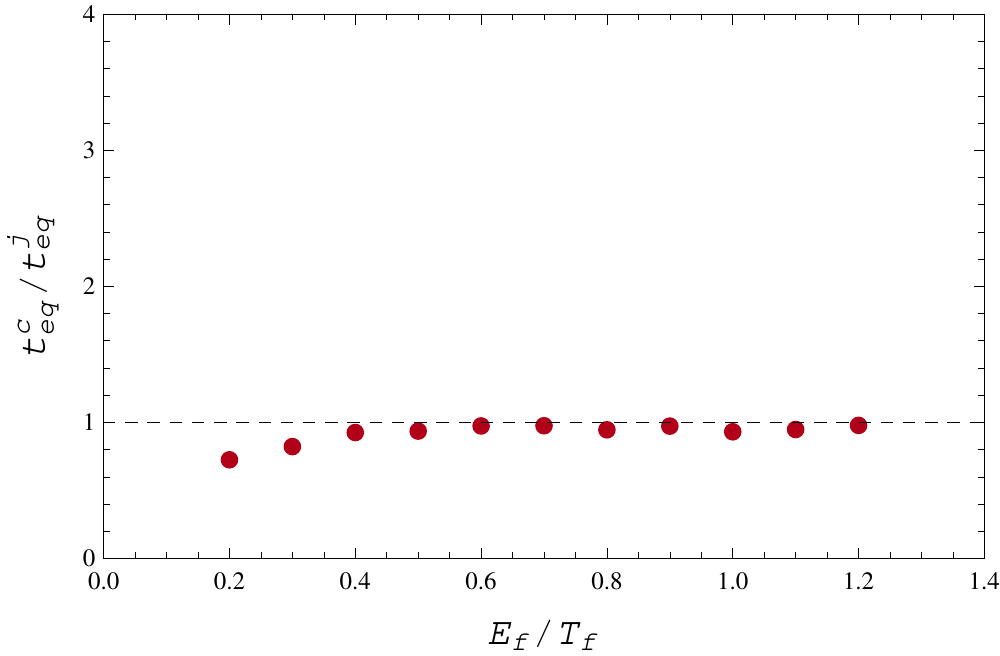}
\hspace{0.6 cm}
\includegraphics[width=8 cm]{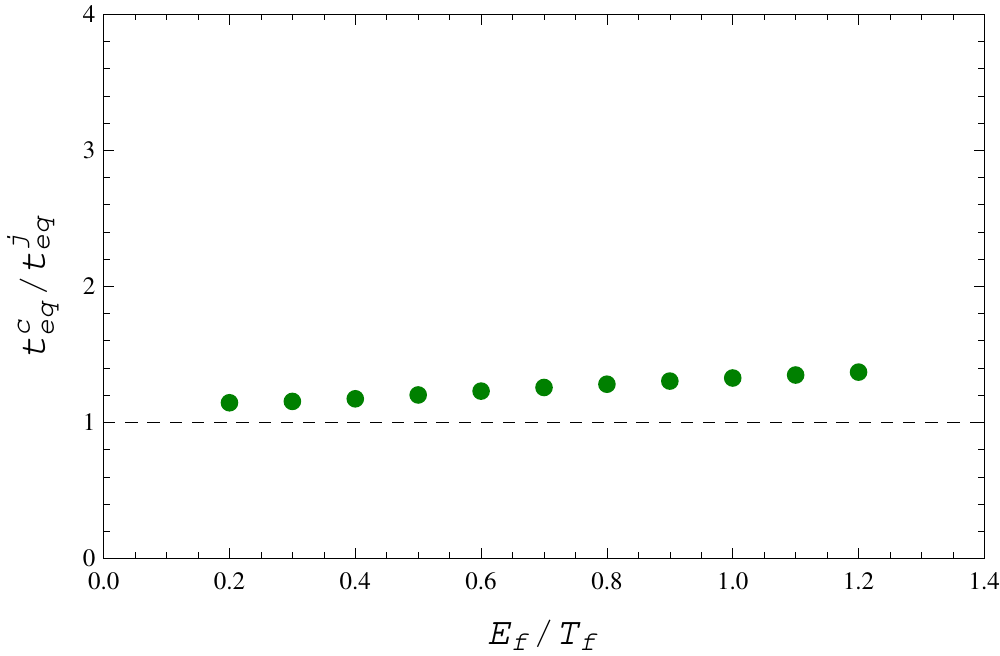}
\caption{$t^c_{eq}/t^j_{eq}$ in terms of $E_f/T_f$ for fast-slow(left) and fast-fast(right) transitions for $T_f=5$. \label{fastfast} }
\end{center}
\end{figure}%

\textbf{\textit{Numerical Results}}: The parameters $k_M$ and $k_E$ are measures to show how fast the functions $M(V)$ and $E(V)$ can reach their maximum values. In other words, $k_M$ and $k_E$ represent the transition times from zero to finite value. It is obvious that for $k_M,k_E\ll 1$ ($k_M,k_E\gg 1$) the transition time is small (large) and hence indicates a slow (fast) growth in the electric field or mass of the black hole. Here we classify our numerical results into four categories as follows
\begin{itemize}
  \item \textit{Slow-Slow transition}\\
  In figure \ref{slowfast}(left), at fixed temperature, $t^c_{eq}/t^j_{eq}$ is plotted in terms of $E_f/T_f$ for $k_M=k_E=5$. In this case the transition times are large enough to consider the system under study near equilibrium. As it is clearly seen from this figure, the equilibration times for the condensation and current are equal, in agreement with the static argument.
  \item \textit{Slow-Fast transition}\\
  The case with $k_M=5$ and $k_E=0.1$ is plotted in figure \ref{slowfast}(right). The equilibration time for condensation is smaller than the corresponding one for the current. It is comprehensible since smaller values of $k_E$ leads to larger deviation from equilibrium \cite{Ali-Akbari:2015gba}.
  \item \textit{Fast-Slow transition}\\
  Now let us consider $k_M=0.1$ and $k_E=5$ plotted in figure \ref{fastfast}(left). In this case for small values of electric field the equilibration time for condensation is smaller than $t_{eq}^j$, that is $t^c_{eq}<t_{eq}^j$. For higher values, it is easy to see that $t_{eq}^c \simeq t_{eq}^j$. It is clearly seen that the ordering of time-scales is in agreement with our preliminary conclusion.
\item \textit{Fast-Fast transition}\\
  Up to now, all observations are consistent with our naive conclusion based on the static case, that is $t_{eq}^c\leq t_{eq}^j$. Now, by considering $k_M=k_E=0.1$, figure \ref{fastfast}(right) indicates that time ordering is reversed, i.e. $t_{eq}^c>t_{eq}^j$. This means that although the current on the D-brane reaches its final equilibrated value, the shape of the brane still changes! Thus, we observe a signal that indicates out-of-equilibrium situation. From this analysis one can conclude that the shape of the probe brane is less important than the electric field in the equilibration time of the current.
\end{itemize}

\begin{figure}[ht]
\begin{center}
\includegraphics[width=8 cm]{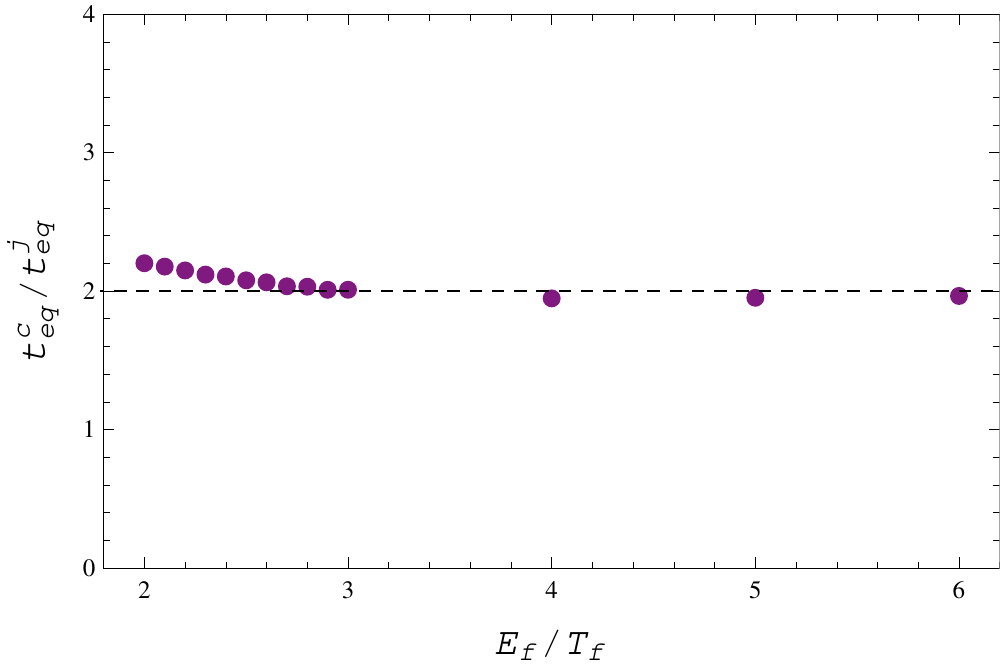}
\hspace{0.6 cm}
\includegraphics[width=8 cm]{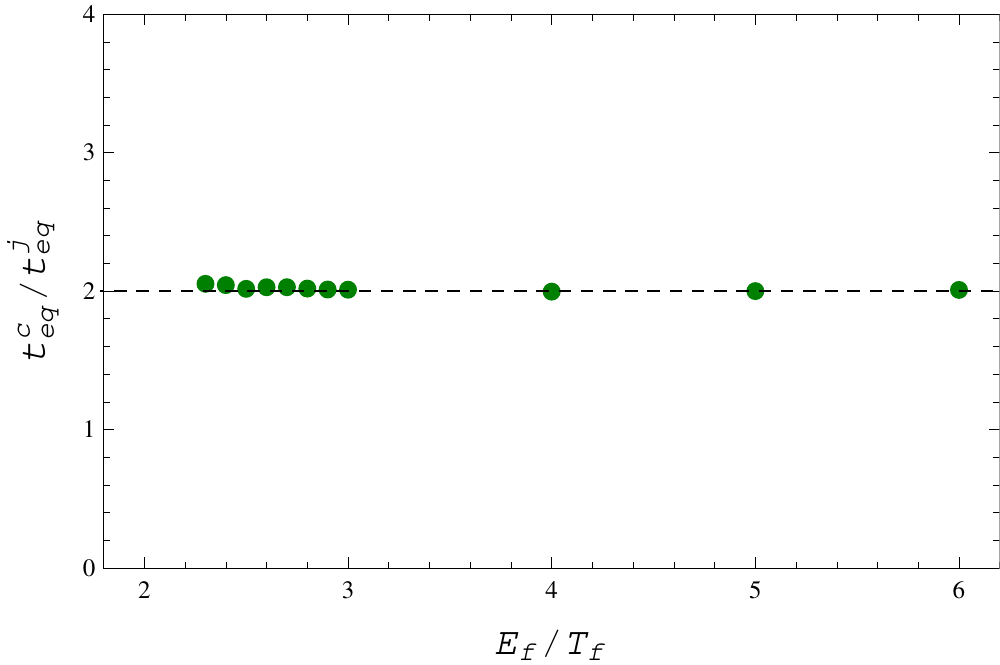}
\caption{$t^c_{eq}/t^j_{eq}$ in terms of $E_f/T_f$ for slow-fast(left) and fast-fast(right) transitions for $T_f=5$. \label{largeE} }
\end{center}
\end{figure}%

\begin{figure}[ht]
\begin{center}
\includegraphics[width=8 cm]{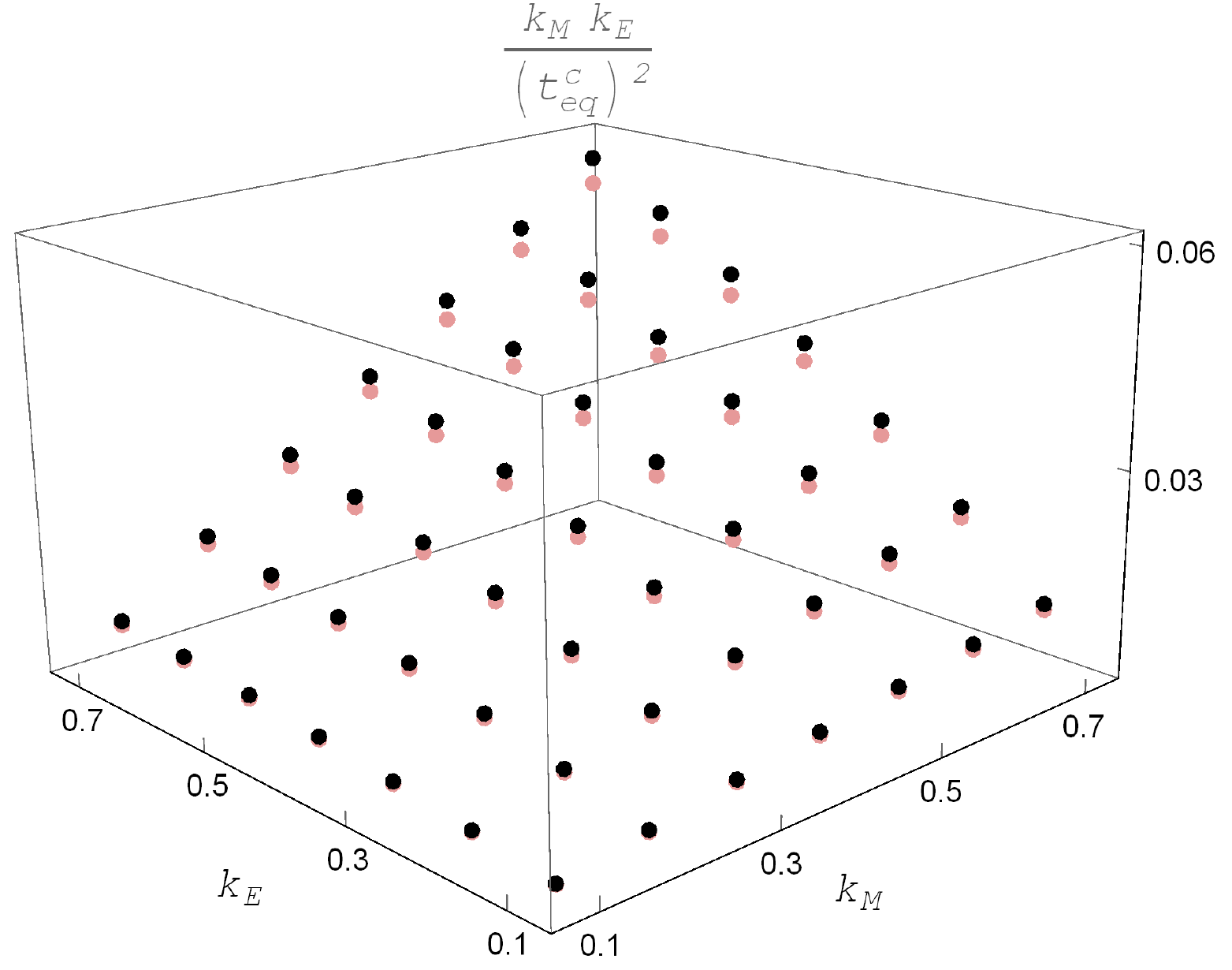}
\hspace{0.6 cm}
\includegraphics[width=8 cm]{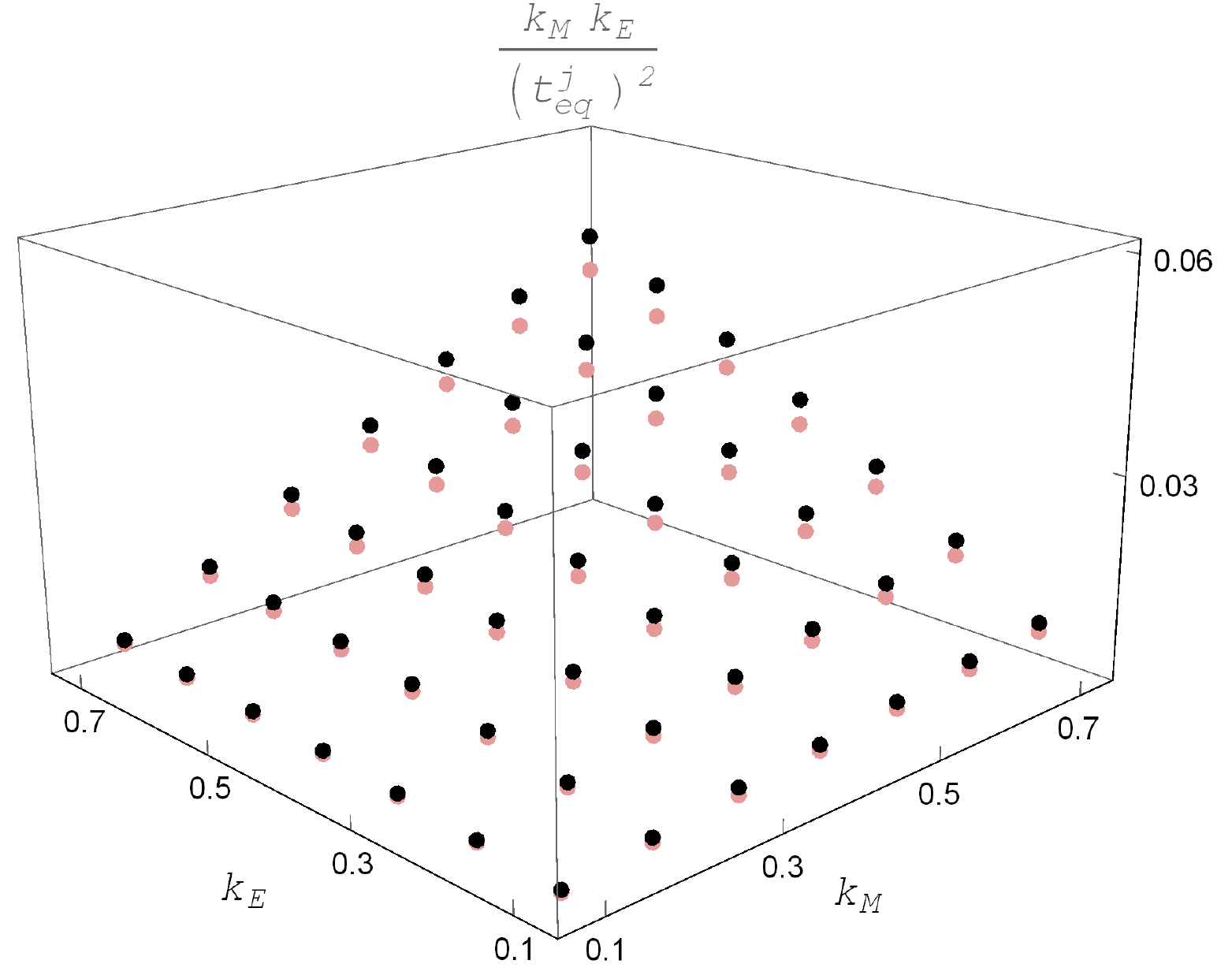}
\caption{The re-scaled equilibration times $k_M k_E (t_{eq}^c)^{-2}$(left) and $k_M k_E (t_{eq}^j)^{-2}$(right) verses $k_M$ and $k_E$ for $T_f=1$ and $E_f=1$ (black) and $E_f=1.1$ (red). \label{threedim} }
\end{center}
\end{figure}%

Now let us consider the slow-fast case in which $E_f/T_f$ is larger than two, figure \ref{largeE}(left). Although in the figure \ref{slowfast}(right) $t_{eq}^c<t_{eq}^j$, figure \ref{largeE}(left) shows that the ratio of the equilibration times is larger than two, that is $t_{eq}^c \gtrsim 2 t_{eq}^j$, meaning that the time ordering of the equilibration times has been reversed.  Therefore, according to the figure \ref{fastfast}(right) and figure \ref{largeE}(left), our numerical results indicate that in order to observe a reversed time ordering a small value of $k_E$ is essential. Then one may guess that the same behaviour can be also observed in figure \ref{fastfast}(left) for large values of $E_f/T_f$ with $k_M=0.1$. But, up to $E_f/T_f \simeq 6$ for which our numerical data is reliable, the same behaviour is not observed. In other words, it seems that a small value of $k_E$ is necessary to reverse time ordering of the equilibration times.

Surprisingly in the figures \ref{largeE} for large value of $E_f/T_f$, the ratio of equilibration times converges to a constant value. But, since the convergence number does depend on the equilibration parameters $\epsilon^c(V)$ and $\epsilon^j(V)$, it does not have a physical interpretation. However, it is important to notice that this convergence is observed by different values of equilibration parameters.  Thus this convergence can be considered as a universal behaviour. Because, for large $E_f/T_f$, the value of $t^c_{eq}/t^j_{eq}$ is almost constant and independent of the final values of the temperature and electric field as it is clearly seen in figures \ref{slowfast} and \ref{largeE}. Therefore we conclude that for large enough $E_f/T_f$ the ratio of the equilibration times behaves universally.

For small values of the transition time, so-called fast quench, another universal behaviour is observed in the gauge theories with holographic dual \cite{Buchel:2013gba}. For example, in \cite{Ali-Akbari:2015gba}, a probe D7-brane in the $AdS_5\times S^5$ background subject to a dynamical external electric field, which is similar to our case, is considered and the calculations show that the re-scaled transition time $kt_{eq}^{-1}$ behaves universally. In this case $t_{eq}$ is equilibration time for the time-dependent current and universality means that $k t_{eq}^{-1}$ is independent of the final value of the electric field. As the case under study in this paper is a generalization of the above system, we expect a universal behaviour as well. However, because there are two transition times $k_M$ and $k_E$, we need to re-scale equilibration times for the condensation and current properly to observe universality. Therefore, in figure \ref{threedim}, we plot the new re-scaled equilibration times $k_M k_E (t_{eq}^c)^{-2}$ and $k_M k_E (t_{eq}^j)^{-2}$ in terms of $k_M$ and $k_E$. Interestingly, at fixed and small value of $k_M$ or $k_E$, a universal behaviour is observed meaning that the re-scaled equilibration times $k_M k_E (t_{eq}^c)^{-2}$ and $k_M k_E (t_{eq}^j)^{-2}$ are independent of the final values of the electric field and temperature. Therefore the smallness of $k_M$ or $k_E$ is enough to observe a universal behaviour. Obviously when both $k_M$ and $k_E$ are small the re-scaled equilibration times behave universally as well.

\section*{Acknowledgment}
M. A. would like to thank School of Physics of Institute
for research in fundamental sciences (IPM) for the research facilities and
environment. The authors would like to thank H. Ebrahim and L. Shahkarami for useful
discussions.

\end{document}